\documentclass[fleqn,usenatbib]{mnras}
\usepackage{newtxtext,newtxmath}
\usepackage[T1]{fontenc}
\usepackage{ae,aecompl}

\usepackage{graphicx}	
\usepackage{amsmath}	
\usepackage{wasysym}    
\usepackage{siunitx}    
\usepackage{longtable}
\usepackage{supertabular}

\usepackage{color}

\definecolor{byzantine}{rgb}{0.74, 0.2, 0.64}

\usepackage[normalem]{ulem}

\newcommand{\bjdtdb}{BJD$_\mathrm{TDB}$}
\defcitealias{Kilkenny1994}{K94}
\defcitealias{Lee2009}{L09}
\defcitealias{Beuermann2012}{B12}

\title[A photometric and dynamical study of HW Vir]{A new photometric and dynamical study of the eclipsing binary star HW Virginis}

\author[S.~B.~Brown-Sevilla et al.]{S. B.~Brown-Sevilla$^{1,3}$\thanks{e-mail address: brown@mpia.de \newline Member of the International Max-Planck Research School for Astronomy and Cosmic Physics at the University of Heidelberg (IMPRS-HD), Germany},
V.~Nascimbeni$^{2,3}$\thanks{e-mail address: valerio.nascimbeni@inaf.it},
L.~Borsato$^{2,3}$,
L.~Tartaglia$^{2}$, \newauthor
D. Nardiello$^{4,2}$,
V.~Granata$^{2,3}$,
M.~Libralato$^{6,2,3}$,
M.~Damasso$^{5}$,
G.~Piotto$^{3,2}$, \newauthor
D. Pollacco$^{7}$,
R. G. West$^{7,8}$,
L.~S.~Colombo$^{2,3}$,
A.~Cunial$^{3,2}$,
G.~Piazza$^{3}$, \newauthor
F.~Scaggiante$^{9}$ \\
\\
$^{1}$Max Planck Institute for Astronomy, K\"onigstuhl 17, 69117, Heidelberg, Germany\\
$^{2}$INAF -- Osservatorio Astronomico di Padova, vicolo dell'Osservatorio 5, 35122 Padova, Italy \\
$^{3}$Dipartimento di Fisica e Astronomia, Universit\`a degli Studi di Padova, Vicolo dell'Osservatorio 3, 35122 Padova, Italy\\
$^{4}$Aix Marseille Univ, CNRS, CNES, LAM, Marseille, France \\
$^{5}$INAF -- Osservatorio Astrofisico di Torino, Via Osservatorio 20, 10025 Pino Torinese, Italy\\
$^{6}$AURA for the European Space Agency (ESA), ESA Office, Space Telescope Science Institute, 3700 San Martin Drive,\\ Baltimore MD 21218, USA\\
$^{7}$Department of Physics, University of Warwick, Gibbet Hill Road, Coventry, CV4 7AL, UK\\
$^{8}$Centre for Exoplanets and Habitability, University of Warwick, Gibbet Hill Road, Coventry CV4 7AL, UK\\
$^{9}$Gruppo Astrofili Salese ``G.~Galilei'', 30036 Santa Maria di Sala (VE), Italy\\
}

\date{Accepted 2021 June 24. Received 2021 June 22; in original form 2021 May 24}

\pubyear{2021}

\begin{document}
\label{firstpage}
\pagerange{\pageref{firstpage}--\pageref{lastpage}}
\maketitle

\begin{abstract}

A growing number of eclipsing binary systems of the ``HW~Vir'' kind (i.~e.,~composed by a subdwarf-B/O primary star and an M dwarf secondary) show variations in their orbital period, also called Eclipse Time Variations (ETVs). Their physical origin is not yet known with certainty: while some ETVs have been claimed to arise from dynamical perturbations due to the presence of circumbinary planetary companions, other authors suggest that the Applegate effect or other unknown stellar mechanisms could be responsible for them.

In this work, we present twenty-eight unpublished high-precision light curves of one of the most controversial of these systems, the prototype HW Virginis. We homogeneously analysed the new eclipse timings together with historical data obtained between 1983 and 2012, demonstrating that the planetary models previously claimed do not fit the new photometric data, besides being dynamically unstable.

In an effort to find a new model able to fit all the available data, we developed a new approach based on a global-search genetic algorithm and eventually found two new distinct families of solutions that fit the observed timings very well, yet dynamically unstable at the $10^5$-year time scale. This serves as a cautionary tale on the existence of formal solutions that apparently explain ETVs but are not physically meaningful, and on the need of carefully testing their stability. On the other hand, our data confirm the presence of an ETV on HW~Vir that known stellar mechanisms are unable to explain, pushing towards further observing and modelling efforts.

\end{abstract}




\section{Introduction}
\label{sec:intro}

\begin{table*}
\caption[HW Vir parameters from the most recent literature]{Orbital and physical parameters of the components of HW Vir from the literature.}
	\centering
	\begin{tabular}{ccccccccccccccccccccc}
	\hline
    \hline
    	Parameter & Primary & & Secondary & Reference \\
        \hline
        Orbital period \textit{P} (days) & & 0.11671967 $\pm$ $\num{1.15e-7}$ & & \cite{Beuermann2012} \\
        Separation \textit{a} ($R_{\astrosun}$) & & 0.860 $\pm$ 0.010 & & \cite{Lee2009} \\
		Inclination \textit{i} ($^{\circ}$) & & 80.98 $\pm$ 0.10 & & \cite{Lee2009} \\
		Eccentricity \textit{e} & & $<$0.0003 & & \cite{Beuermann2012} \\
		Distance \textit{d} (pc) & & 181 $\pm$ 20 & & \cite{Lee2009} \\
		Mass ($M_{\astrosun}$) & 0.485 $\pm$ 0.013 & & 0.142 $\pm$ 0.004 & \cite{Lee2009} \\
		Radius ($R_{\astrosun}$) & 0.183 $\pm$ 0.026 & & 0.175 $\pm$ 0.026 & \cite{Lee2009} \\
        Temperature (K) & 28488 $\pm$ 208 & & 3084 $\pm$ 889 & \cite{WoodSaffer1999} and \\
        & & & & \cite{Lee2009} \\
        Visual magnitude ($V$ band) &  & 10.6 (combined) & --- & \citet{Zacharias2012}  \\
        Bolometric magnitude $M_{\textrm{bol}}$ (mag) & 1.46 $\pm$ 0.24 & & 11.20 $\pm$ 0.46 & \cite{Lee2009} \\
        Absolute Visual magnitude $M_V$ (mag) & 4.22 $\pm$ 0.24 & & 15.59 $\pm$ 0.46 & \cite{Lee2009} \\
        Bolometric luminosity $L_{\mathrm{bol}}$ ($L_{\astrosun}$) & 19.7 $\pm$ 5.6 & & 0.003 $\pm$ 0.001 & \cite{Lee2009} \\
        \hline
	\end{tabular}
\label{tab:hwv}
\end{table*}

The discovery of the first exoplanets by \citet{Wolszczan1992} and \cite{MayorQueloz1995} was the starting point to the detection of a great number of other planetary systems through different observing techniques. Although the majority of them have been found orbiting Sun-like stars \citep[e.g.,][]{Petigura2015}, there is an increasing number of exoplanets being discovered orbiting all kinds of stars \citep[e.g.,][]{Gould2014, Gillon2017, Brewer2018}. A particularly interesting case among them is represented by circumbinary planets, which orbit a binary system instead of a single star. This kind of planets can be detected, among other techniques (such as transits, e.g., \cite{Kostov2016}; radial velocity, e.g., \cite{Konacki2009}; or light travel-time delay, e.g., \cite{Silvotti2018}), by measuring and analysing changes in the orbital period of eclipsing binary stars, a dynamical method commonly known as Eclipse Time Variations \citep[ETV, e.g.][]{Sale2020}. These variations have been observed in a wide range of binary systems, such as post-common envelope binaries, for example, which exhibit modulation periods of a few tens of years \citep[e.g.][]{Bours2016}. A possible mechanism to explain ETVs is the light travel time effect (LTTE; also known as R\o mer effect), which refers to the combination of the motion of the stellar components with respect to the barycenter of the system due to the gravitational perturbation of additional bodies, with the finite speed of light \citep{Irwin1952}.

Among the vast taxonomy of eclipsing binaries, the so called ``HW Virginis" (HW Vir) systems have recently drawn the attention of astronomers. These systems are post-common envelope binaries composed of a sub-dwarf of spectral type O or B and a late-type main sequence star, e.g.~sdB+dM for the prototype.
They have very short orbital periods (of the order of a few hours), and in a surprisingly high fraction of the cases, ETVs have been observed, typically from tens of seconds to several minutes of amplitude and semi-regular modulations on long time scales, from years to decades (see \citealt{Heber2016} for a detailed review on HW Vir systems).
Different explanations have been proposed to interpret ETVs, usually based on two different effects or a combination of them:
the LTTE effect caused by one or more unseen companions, and the so called Applegate effect. The latter was first proposed by \cite{Applegate1992}, and it interprets the variations on the orbital period as a consequence of magnetic activity in one of the stars of the binary system (in the case of HW Vir the main-sequence component). According to \citet{Applegate1992}, the distribution of the angular momentum in the active star changes as the star goes through its activity cycle. These variations on the angular momentum distribution induce a change in the gravitational quadrupole moment of the star (making it more or less oblate), which can cause perturbations in the orbit of the system and thus in the orbital period.

In this work we analyse data from the prototypical HW Vir, a detached eclipsing binary system first identified as such by \cite{MenziesMarang1986}. HW Vir has a very short period of 2.8~h, and its components have masses of 0.49 and 0.14 $M_{\astrosun}$, for the sdB and dM components, respectively (see Table \ref{tab:hwv} for the most recent parameters of HW Vir). Since its discovery, the system has been broadly studied due to its intrinsic characteristics and its striking period variations. A decrease in the orbital period of the system was first detected by \cite{Kilkenny1994}, followed by \cite{Cakirli1999} who re-analysed the eclipse timings between 1984-1999 and concluded that LTTE was the most promising explanation for the observed period variations. They proposed that HW Vir was revolving about a third body with a period of 19 years.
Later on, further studies were performed \citep{WoodSaffer1999, Kilkenny2000, Kiss200} analysing the period variations with different techniques, without reaching a definitive explanation.
\cite{Kilkenny2003} presented new eclipse timings for HW Vir and confirmed the presence of a periodic LTTE term due to a third body in the system, a claim also supported by \cite{Ibanoglu2004}.

\cite{Lee2009} presented new CCD photometry with a 8-year baseline,
and proposed that the linear term of the period decrease ($\mathrm{d}P/\mathrm{d}t$) may be caused by angular momentum loss due to magnetic stellar wind braking, while the cyclic period variations may be interpreted as LTTE terms induced by the presence of two additional bodies in the system, having masses of $M_3\sin{i_3}= 19.2$~$M_\mathrm{J}$ and $M_4\sin{i_4}= 8.5$~$M_\mathrm{J}$, respectively\footnote{$\sin i_3$ and $\sin i_4$ being the inclination with respect to the line of sight of the orbital plane of the inner and outer perturber, respectively. Throughout this paper we adopt this index convention, meaning the third and fourth massive bodies of the system.}.
This model was independently tested by \cite{Beuermann2012}, who found that it fails to fit their new eclipse timings and it is dynamically unstable on a time scale of a few thousand years.
\citeauthor{Beuermann2012} also proposed a new LTTE model with two companions with masses $M_3\sin{i_3} \simeq 14$~$M_\mathrm{J}$ and $M_4\sin{i_4}= 30$-$120$~$M_\mathrm{J}$, and periods of 12.7 yr and $55\pm{15}$ yr, respectively. \cite{Horner2012} independently tested \citeauthor{Lee2009}'s model and came to the same conclusion about the dynamical instability of the system on very short timescales; they also claimed that the ETVs cannot be driven by gravitational influence of perturbing planets only, and that there must be another astrophysical mechanism taking place in order to explain them.

Finally, \cite{Esmer2021}; found a new two-planet solution, but it did not appear to be dynamically stable. The main differences between our approach and theirs will be summarized in the Discussion.

Regarding the Applegate effect, \cite{Navarrete2018} analysed the required energy to drive the Applegate effect in a sample of 12 close binary systems (including HW Vir), and compared it with the energy production of a simulated sample of magnetically active stars. In the case of HW Vir, they discarded the possibility of this effect being the underlying cause for the ETVs, since the magnetic field of the magnetically active star (i.e. the dM star) is not strong enough to produce these variations.

A conclusive explanation to HW Vir's ETVs is still missing. For this reason, our aim is to derive new eclipse timings from our unpublished photometric data, and use them along with the ones available in the literature to better constrain the physical parameters characterising the system of HW Vir, as well as to test these new parameters for dynamical stability on a large timescale.

The paper is organized as follows:
in Section~\ref{sec:obs} we present our data, along with the data reduction process we followed, the light curve fitting, and the determination of the eclipse timings, while in Section~\ref{sec:met} we outline the LTTE modelling and test the previous model proposed to explain the ETVs of the system with our new data, as well as using an N-body integrator to test its dynamical stability. In Section~\ref{sec:new} we describe the method we used to estimate new parameters for the putative companions of HW Vir. In Section~\ref{sec:con} we discuss our findings and we draw some conclusions regarding the explanation behind the ETVs of HW Vir as well as some prospects for future work.

\begin{table*}
\caption{Log of observations. The columns give: a unique identifier (matching those in Fig.~\ref{fig:lc}), the ``evening date'' of the observation, the telescope used, the number of acquired frames, the photometric passband, and, which eclipses were observed among the primary and secondary.}
	\begin{tabular}{cccrcccr}
	\hline
    \hline
	ID & ``Evening'' date & Telescope & $N_\textrm{frames}$ & Filter & Phase coverage \\
    \hline
    \texttt{w1-w4}&2008-2012 & WASP-South & 18\,410 & WASP (clear) & Both (multiple)\\
    \texttt{s1}&2012/03/11 & Asiago Schmidt & 321 & $R$-Bessel & Both\rule{0pt}{15pt} \\
    \texttt{s2}&2012/03/12 & Asiago Schmidt & 332 & $R$-Bessel & Both \\
    \texttt{s3}&2018/04/20 & Asiago Schmidt & 557 & $r$-Sloan & Primary \\
    \texttt{g1}&2014/03/12 & GAS & 280 & $V$-Bessel & Both\rule{0pt}{15pt} \\
    \texttt{g2}&2014/03/28 & GAS & 728 & $V$-Bessel & Both twice \\
    \texttt{g3}&2014/03/29 & GAS & 660 & $V$-Bessel & Both twice \\
    \texttt{g4}&2014/03/30 & GAS & 304 & $V$-Bessel & Primary and partial secondary \\
    \texttt{g5}&2014/03/31 & GAS & 700 & $V$-Bessel & Both twice \\
    \texttt{g6}&2014/05/24 & GAS & 325 & $V$-Bessel & Both \\
    \texttt{c1}&2011/02/05 & Asiago 1.82-m & 326 & $R$-Bessel & Partial primary\rule{0pt}{15pt} \\
	\texttt{c2}&2012/01/26 & Asiago 1.82-m & 1\,392 & $V$-Bessel & Both \\
	\texttt{c3}&2013/02/04 & Asiago 1.82-m & 448 & $V$-Bessel & Primary \\
	\texttt{c4}&2013/02/07 & Asiago 1.82-m & 929 & $V$-Bessel & Both \\
	\texttt{c5}&2014/03/06 & Asiago 1.82-m & 1\,252 & $V$-Bessel & Primary \\
	\texttt{c6}&2014/04/01 & Asiago 1.82-m & 1\,086 & $V$-Bessel & Primary \\
	\texttt{c7}&2015/03/13 & Asiago 1.82-m & 320 & $r$-Sloan & Partial primary \\
	\texttt{c8}&2016/02/05 & Asiago 1.82-m & 620 & $V$-Bessel & Primary \\
    \texttt{c9}&2016/02/08 & Asiago 1.82-m & 1\,122 & $V$-Bessel & Both \\
	\texttt{c10}&2017/01/21 & Asiago 1.82-m & 1\,943 & $r$-Sloan & Primary \\
	\texttt{c11}&2017/02/25 & Asiago 1.82-m & 1\,663 & $r$-Sloan & Both \\
	\texttt{c12}&2017/03/02 & Asiago 1.82-m & 950 & $r$-Sloan & Primary \\
	\texttt{c13}&2019/01/03 & Asiago 1.82-m & 1\,632 &$r$-Sloan & Both \\
	\texttt{c14}&2019/03/12 & Asiago 1.82-m & 713 &$r$-Sloan & Primary \\
	\texttt{c15}&2019/03/31 & Asiago 1.82-m & 1\,170 &$r$-Sloan & Both \\
    \texttt{kt1-2}&2016      & K2 & 89\,970 & K2 (clear) & Both (multiple)\rule{0pt}{15pt} \\
    \hline
	\end{tabular}
\label{tab:obs}
\end{table*}

\section{Observations and data reduction}
\label{sec:obs}

Our analyzed data set consists of thirty photometric observations of HW Vir obtained in a timespan of $\sim$11 years, (2008 to 2019), including twenty-eight previously unpublished light curves.
For our analysis we combined data from five different instruments as described in the following.

From the Asiago Astrophysical Observatory located on Mt. Ekar in Asiago, we obtained fifteen light curves using the $1.82\,\rm{m}$ ``Copernico" telescope and the Asiago Faint Object Spectrograph and Camera (AFOSC). These images were taken with an exposure time ranging from 2 to 6~s, through the $V$, $R$ and $r$ filters.
Three light curves were obtained using the $67/92\,\rm{cm}$ Schmidt telescope located at the same observatory. These observations were carried out in the $R$ and $r$ filters with an exposure time of 20~s, except for the last one (4~s).

Six light curves were obtained using the telescopes of the ``Gruppo Astrofili Salese Galileo Galilei"\footnote{\url{https://www.astrosalese.it/}}, the telescopes have a primary mirror of $410$~mm of diameter and a focal length of $1710$~mm and they are located in Santa Maria di Sala, in northern Italy. The observations were carried out in the $V$ filter and with an exposure time ranging from 20~s to 45~s.

Our largest set in terms of number of data points comes from WASP-South, a transit survey with an array of small telescopes operating at SAAO in South Africa \citep{Pollacco2006}. WASP-South gathered four full seasons of observations of HW Vir from 2008 to 2012, for a grand total of 353 measured primary eclipses. This particular data set has not yet been included in a public data release, and has been kindly provided to us by the WASP-South team.

We also include in our analysis two light curves from $K2$ \citep{Howell2014}, observed during Campaign 10, and a vast collection of literature timings already analysed by \citet{Beuermann2012} and summarized at the end of this Section.
A detailed summary of all the observations is given in Table~\ref{tab:obs}. Each light curve is identified with a unique ID with the leading letter matching the telescope: \texttt{w} for WASP-South, \texttt{s} for Asiago Schmidt, \texttt{g} for GAS, \texttt{c} for Asiago Copernico, \texttt{kt} for K2. The \texttt{w} and \texttt{kt} light curves are split in four and two ``chunks'' (respectively), for the reasons explained in Section \ref{eclipsetiming}.

\begin{figure*}
	\centering
        \includegraphics[width=0.8\textwidth]{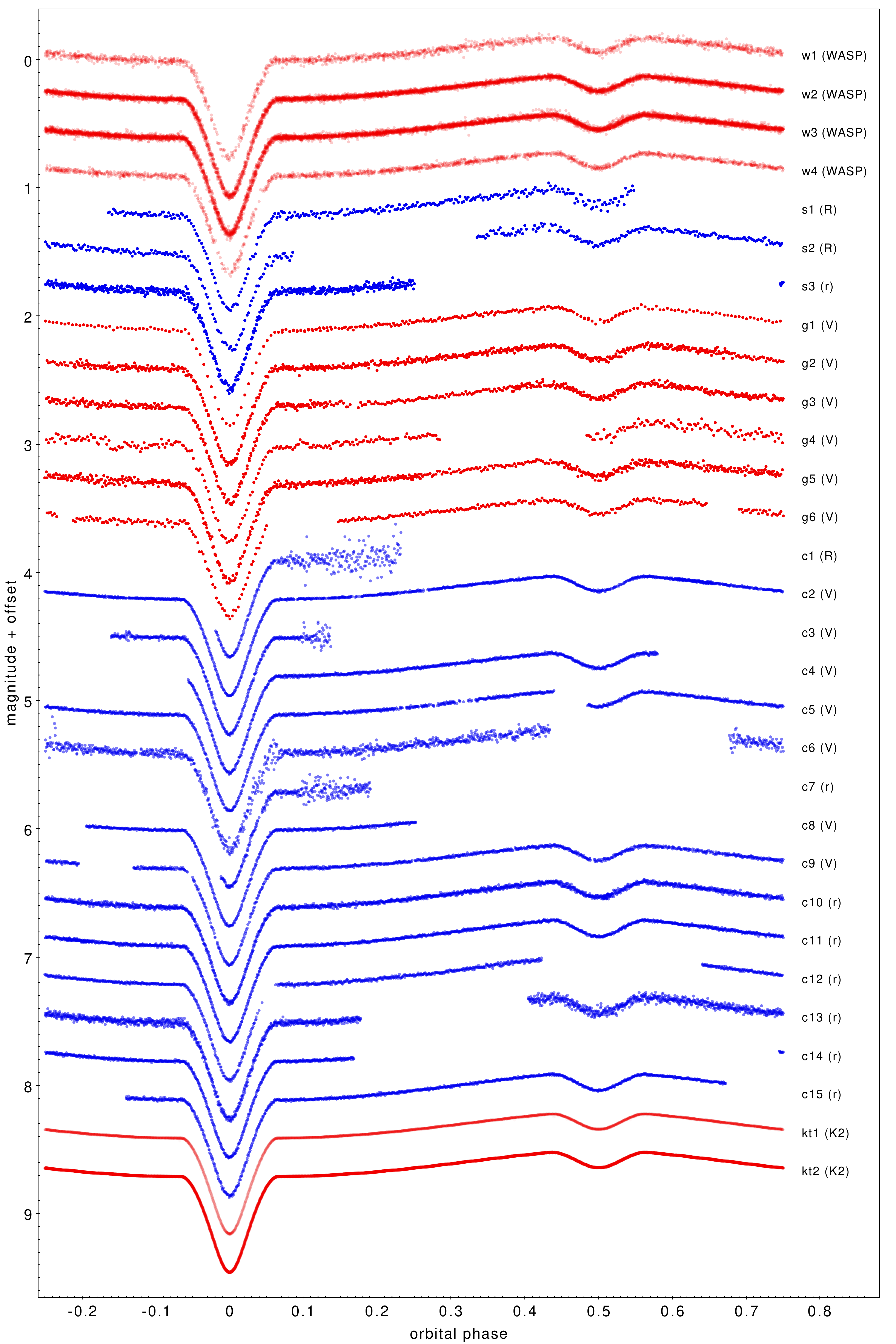}
	\caption[Light curves of HW Vir]{The thirty light curves of HW Virginis analyzed in the present study, plotted as a function of the orbital phase. Each curve is labeled with an identifier (matching those in Table \ref{tab:obs}) and the filter name (uppercase for the Bessel system, lowercase for SDSS). The SuperWASP (\texttt{w1-w4}) and $K2$ (\texttt{kt1-kt2}) curves are split into separate chunks as described in Section \ref{eclipsetiming}. The color scheme is used for visual reference to identify each set of light curves.}
	\label{fig:lc}
\end{figure*}

Due to the lack of stellar crowding in the field of HW Vir, we use the differential aperture photometry technique to reduce our photometric series from the \texttt{c}, \texttt{s} and \texttt{g} data sets. To perform the usual data reduction and the aperture photometry we use the software \texttt{STARSKY}, a pipeline written in \texttt{Fortran 77/90} by \citet{Nascimbeni2011, Nascimbeni2013}, that was specially developed for The Asiago Search for Transit timing variations of Exoplanets (TASTE) project.
As for the \texttt{w} data set, we take the light curves as they were delivered by the standard WASP software pipeline.
For the $K2$ data, we extracted the light curve by reconstructing the 89\,970 images containing HW Vir as done in \cite{Libralato2016}, and performing a 3-pixel aperture photometry of the target on each image, subtracting to the total flux the local background measured in an annulus centred on the target and having radii $r_{\rm in}=7$ pixels and $r_{\rm out}=15$ pixels. We detrended the light curve following the procedure by \cite{Nardiello2016}. The resulting light curves from all the observations are shown in Fig.~\ref{fig:lc}.

In order to measure timing variations with an absolute accuracy much better than one minute, as needed for measuring ETVs, it is crucial to convert all our time stamps to a single, uniform time standard. Therefore we convert all of them to the so called Barycentric Julian Date computed from the Barycentric Dynamical Time,
or BJD$_\mathrm{TDB}$, following the prescription by \citet{Eastman2010}. For this task we rely on the \texttt{VARTOOLS} code\footnote{\url{https://www.astro.princeton.edu/~jhartman/vartools}}. Due to the crucial importance of this step for our dynamical analysis, we perform a double-check of the conversion with the help of the on-line tool\footnote{\url{http://astroutils.astronomy.ohio-state.edu/time/}} made available by \citet{Eastman2010}.
We also apply this time conversion to all the 287 literature timings from SAAO, \cite{Wood1993}, \cite{Lee2009}, BAV, VSNET, AAVSO, BRNO and \cite{Beuermann2012}, who, in turn, used timings from MONET/North. Again, all the HJD$_\mathrm{UTC}$ and BJD$_\mathrm{UTC}$ are homogeneously converted to BJD$_\mathrm{TDB}$ to ensure a proper comparison between the old timings and our new ones. A comprehensive listing of all the literature timings as converted by us is available in Appendix A.

\section{Eclipse timings}\label{eclipsetiming}

To retrieve the best estimate of the orbital and physical parameters of the system, and most crucially the eclipse central time $T_0$, we fit an appropriate model to our light curves.
For this purpose we use the \texttt{JKTEBOP}\footnote{\url{http://www.astro.keele.ac.uk/jkt/codes/jktebop.html}} code \citep{Southworth2012}, which was originally developed to fit light curves of detached eclipsing binaries and later adapted to model also exoplanetary transits.
\texttt{JKTEBOP} implements non-linear least-squares optimisation techniques
\citep[based on the Levenberg-Marquardt algorithm,][]{More1978}. It has different ``tasks'' to choose from, according to how the light curves would be fitted and how the uncertainties are estimated.
This process is meant to determine the best-fitting values of $T_0$ for each individual light curve and a reliable error estimate.

As a first step, we check that the software is properly fitting our light curves and converging to a physical solution by using \texttt{task3}, i.e., by simply running the task to each preliminary light curve and performing a visual inspection.
At this stage, we decide to split the \texttt{w} and \texttt{kt} light curves in separate ``chunks''. For the WASP-South data, this is done because the composite light curve has a four-year coverage, and fitting it as a whole could in principle smear the LTTE signal; by splitting it into four distinct ``seasons'' of about four months each we completely avoid this risk (the shortest significant $O-C$ periodicity reported in the literature being $\sim$3000~days). As for the $K2$ data, the Campaign 10 light curve shows a large two-week gap due to a repointing procedure followed by an unexpected shutdown of the camera. To make ourselves sure that there are no systematic errors introduced by this issue, we separately analyzed the two chunks before and after the blank gap.

We then remove the outliers from our light curves at $4\sigma$
using \texttt{task4} of \texttt{JKTEBOP}, and, since we want to obtain a reliable measure of the eclipse time ($T_0$), we need to first build consistent templates of the parameters for each of the filters of our observations, to leave only $T_0$ as a free parameter in the final fit. To do this, we join the full-phase light curves from the same filter (since the light curves are colour dependent) and leave the following parameters free to find the best-fitting values:
the sum of the stellar radii $R_1+R_2$, their ratio $R_1/R_2$, the inclination of their orbit, the surface brightness and the limb darkening of the primary star, the reflection coefficient of the secondary star, the scale factor and the eclipse time ($T_0$).
We do this for the $V$ and $R/r$ filters, and additionally, for the WASP and $K2$ light curves. Then we run \texttt{task9} of \texttt{JKTEBOP}, which uses a \textit{residual-shift} method to obtain the best fit.
This method evaluates the best fit for the data points and shifts the residuals of the fit point-by-point through all the data, calculating a new best fit after each shift. This approach allows to have as many best fits as points in the input light curve, and it also estimates the relevance of the correlated red noise to the parameters of the fit. The output of this task are therefore three high-accuracy parameter sets (templates), one for each filter:
a $V$ template for the Copernico/$V$ and GAS light curves;
an $R/r$ template for the Copernico/$R,r$ and the Schmidt/$R$ light curves; and an unfiltered template for the WASP and the $K2$ light curves.

\begin{figure*}
        \includegraphics[width=0.9\textwidth]{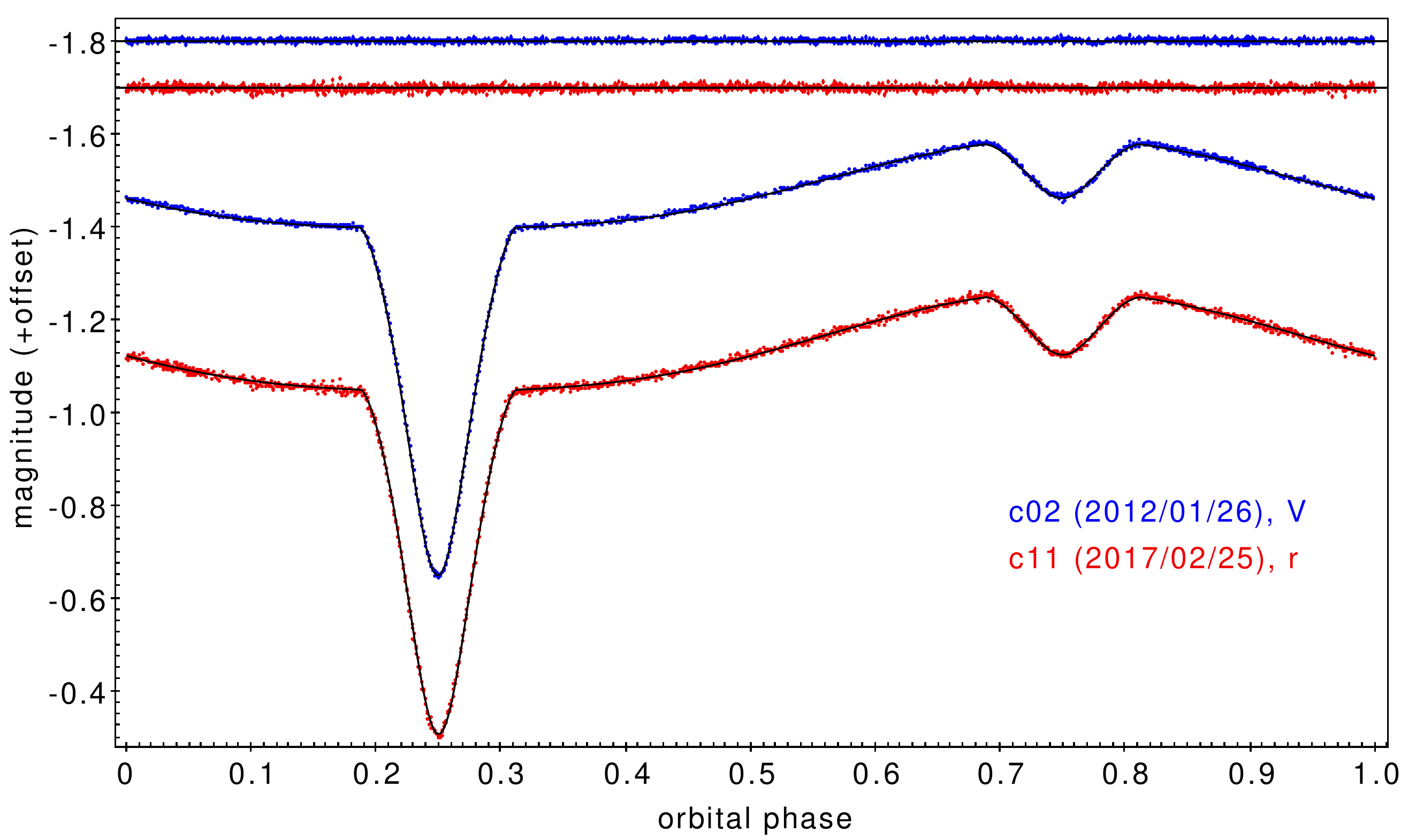}
	\caption{JKTEBOP best fit models on our two most complete light curves from the Copernico telescope: $\texttt{c2}$ in Bessel $V$ (blue points), and $\texttt{c11}$ in Sloan $r$ (red points). The residuals are shown in the upper part of the plot; their rms scatter is 3.1 and 5.0 mmag, respectively.}
	\label{fig:lcfit}
\end{figure*}

We retrieve the $T_0$s by running \texttt{task9} one more time fixing all the parameters except the eclipse times.
An example of the quality of the fit on our two most complete light curves from the Copernico telescope ($\texttt{c2}$ in Bessel $V$, and $\texttt{c11}$ in Sloan $r$) is shown in Fig.~\ref{fig:lcfit}.

The resulting timings of HW Vir are reported in Table \ref{tab:t0}.
We compute a total of 30 mid-eclipse timings, with an excellent median timing error for our light curves of only $\sim$1.3~s and down to 0.3~s for the best ones (from the \texttt{c} and \texttt{w} sets). Our new data increases the current number of high-precision observations ($\sigma(T_0)<5$~ s) by about 50\%, and extends the baseline by six years with respect to the dynamical study of HW Vir \citep{Beuermann2012}.

\begin{table}
\caption[Eclipse Timings of HW Vir obtained with our data]{Best-fitting eclipse timings ($T_0$) for the primary eclipse of HW Vir derived from our unpublished data. The epoch is computed with respect to the linear ephemeris in Eq.~\ref{eqn:TcB}.}
	\centering
	\begin{tabular}{llrc}
	\hline
    \hline
    $T_0$ (BJD$_\mathrm{TDB}$) & $\sigma_{T_0}$ (days) & Epoch & ID \\
        \hline
        2455598.608756 & 0.000039 & 468 & \texttt{c1}\rule{0pt}{12pt} \\
        2455953.669686 & 0.000004 & 3510 & \texttt{c2} \\
        2456328.572882 & 0.000004 & 6722 & \texttt{c3} \\
        2456331.607585 & 0.000007 & 6748 & \texttt{c4} \\
        2456723.551785 & 0.000006 & 10106 & \texttt{c5} \\
		2456749.463503 & 0.000022 & 10328 & \texttt{c6} \\
        2457095.536914 & 0.000035 & 13293 & \texttt{c7} \\
        2457424.685857 & 0.000003 & 16113 & \texttt{c8} \\
        2457427.603842 & 0.000006 & 16138 & \texttt{c9} \\
        2457775.661358 & 0.000010 & 19120 & \texttt{c10} \\
        2457810.560486 & 0.000007 & 19419 & \texttt{c11} \\
        2457815.579418 & 0.000009 & 19462 & \texttt{c12} \\
        2458487.650169 & 0.000014 & 25220 & \texttt{c13} \\
        2458555.580938 & 0.000003 & 25802 & \texttt{c14} \\
        2458574.372776 & 0.000008 & 25963 & \texttt{c15} \\
        2455998.606687 & 0.000022 & 3895 & \texttt{s1}\rule{0pt}{12pt}\\
        2455999.657099 & 0.000048 & 3904 & \texttt{s2} \\
        2458229.466702 & 0.000026 & 23008 & \texttt{s3} \\
        2456729.504448 & 0.000022 & 10157 & \texttt{g1}\rule{0pt}{12pt}\\
        2456745.495025 & 0.000015 & 10294 & \texttt{g2} \\
        2456746.428763 & 0.000023 & 10302 & \texttt{g3} \\
        2456747.479294 & 0.000050 & 10311 & \texttt{g4} \\
        2456748.413045 & 0.000025 & 10319 & \texttt{g5} \\
        2456802.454163 & 0.000046 & 10782 & \texttt{g6} \\
        2454539.612655 & 0.000012 & $-$8605 & \texttt{w1}\rule{0pt}{12pt}\\
		2454961.436853 & 0.000003 & $-$4991 & \texttt{w2} \\
        2455283.582736 & 0.000004 & $-$2231 & \texttt{w3} \\
        2455596.741284 & 0.000008 & 452 & \texttt{w4} \\
        2457584.4748480 & 0.0000003 & 17482 & \texttt{kt1}\rule{0pt}{12pt}\\
        2457629.1784108 & 0.0000002 & 17865 & \texttt{kt2} \\
        \hline
	\end{tabular}
\label{tab:t0}
\end{table}

\begin{figure}
	\centering
		\includegraphics[width=0.45\textwidth]{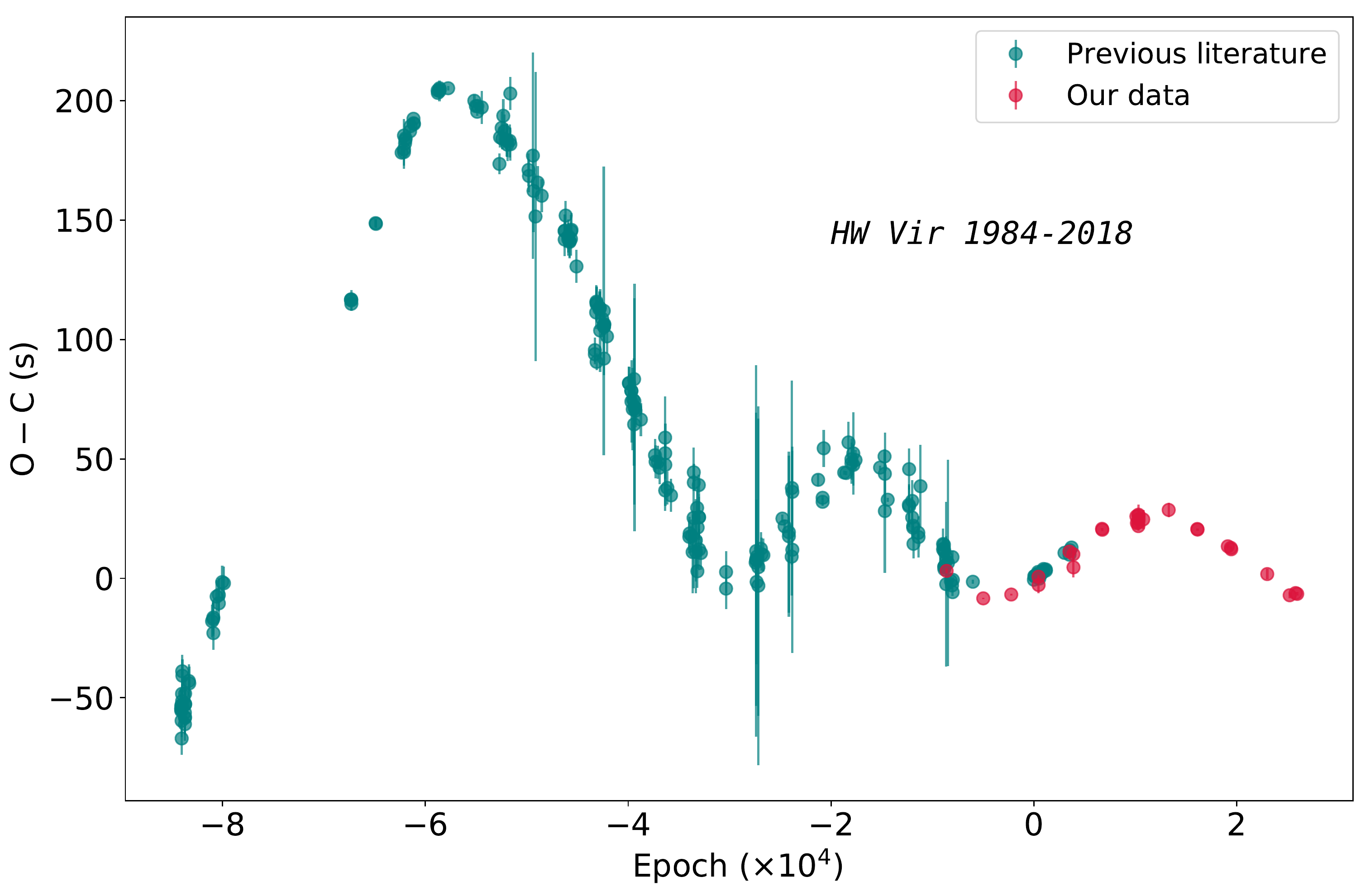}
	\caption[O-C diagram of HW Vir with the literature values plus our data]{Observed$-$calculated T$_{0}$ diagram of HW Vir built with the literature data plus our data. We use Eq.\ref{eqn:TcB} to obtain the linear ephemeris ($T_\mathrm{c}$, see text) and compute the \textit{O-C}.} 
	\label{fig:o-clit}
\end{figure}

We build the observed minus calculated ($O-C$) diagram for HW Vir by plotting both the new and the old eclipse timings as a function of the epoch \textit{E}, using the linear ephemeris formula derived by \citeauthor{Beuermann2012}, by fitting their mid-eclipse times alone

\begin{equation}
T_\mathrm{c} = 2\,455\,543.984055(2)+0.116719555(2)\times E,
\label{eqn:TcB}
\end{equation}

\noindent where $T_\mathrm{c}$ is the calculated time of the primary eclipse in the $\textrm{BJD}_\textrm{TDB}$ time standard.
In Fig. \ref{fig:o-clit} we show the \textit{O-C} diagram including all the up-to-date eclipse timings of HW Vir.
As it can be seen, our new data match the existing one with a remarkable precision (within 1$\sigma$), which also serves as an external check for our absolute time calibration.

\section{Modelling}
\label{sec:met}

\subsection{LTTE calculation}
\label{sec:ltte}

To calculate the LTTE we develop a \texttt{Fortran 77} code that implements an adaptation of the equation by \cite{Irwin1952} to compute the LTTE:

\begin{equation}
\tau_k = K_k\Biggl[\frac{1-e_k^2}{1+e_k\cos v_k}\sin(v_k+\omega_k)\Biggr]\textrm{ ,}
\label{eqn:ltt}
\end{equation}

\noindent where the subindex $k=1,2,...$ indicates the stellar or substellar companion causing the modulation, $\tau_k$ is the light-time delay, $e_k$ is the eccentricity of the orbit, $\omega_k$ is the argument of periastron, $\nu_k$ is the true anomaly, and $K_k$ is the semi-amplitude of the modulation given by

\begin{equation}
K_k = \frac{a_{k,\mathrm{bin}} \sin i_k}{c}\textrm{ ,}
\label{eqn:K}
\end{equation}

\noindent where $a_{k,\mathrm{bin}}$ is the semi-major axis of the orbit of the binary around the common centre of mass, $i_k$ is the inclination of the orbit with respect to the line of sight,
and $c$ is the speed of light.

The approach of \cite{Irwin1952}, was to use the plane perpendicular to the line of sight that passes through the centre of the elliptical orbit of the binary about the centre of mass of all the bodies in the system as the reference frame, which adds a $e_k\sin \omega_k$ term to Eq. \ref{eqn:ltt}.
Our approach is to use another perpendicular (and parallel) plane
to the line of sight that passes through the centre of mass of all the bodies in the system as the reference frame, resulting in the exclusion of this term.

\subsection{Test of the previous model}

By fitting a model with the contribution of two LTTE terms ($\tau$, described in Section~\ref{sec:intro}), \citet{Beuermann2012} derived an underlying linear ephemeris for the binary given by

\begin{equation}
T_\mathrm{c} = 2\,455\,730.550186(3)+0.116719675(6)\times E.
\label{eqn:TcBm}
\end{equation}

\begin{figure}
	\centering
		\includegraphics[width=0.48\textwidth]{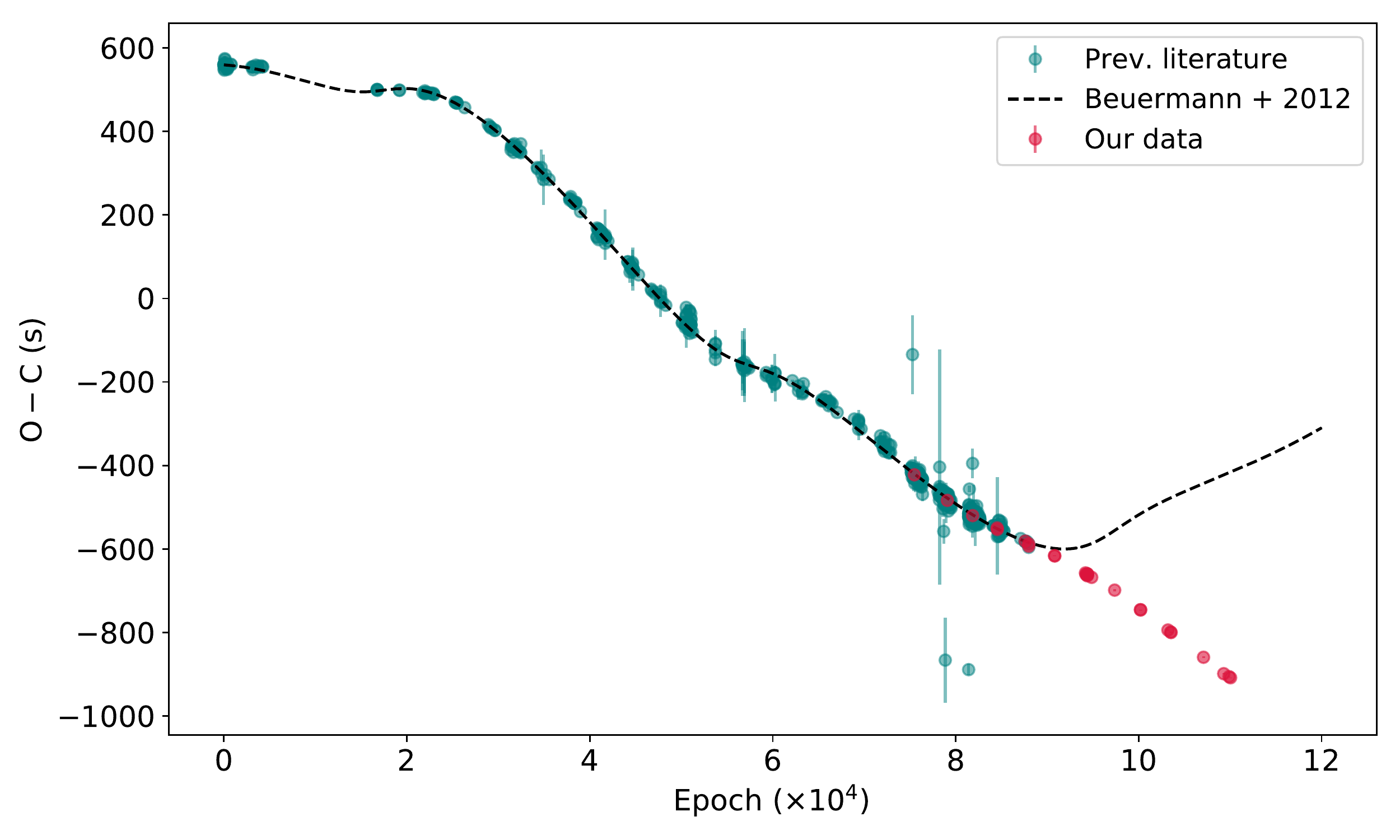}
	\caption[Literature model and data with our data]{\textit{O-C} diagram of HW Vir showing \citeauthor{Beuermann2012}'s model along with all the literature timings available, with the model extended along time and our new timings over-plotted for comparison. Some of the error bars fall within the size of the points. We use Eq.~\ref{eqn:TcBm} to obtain the linear ephemeris and to compute the the \textit{O-C}.}
	\label{fig:ocl}
\end{figure}

To test this two-companion model, we plot the \textit{O-C} diagram using both the literature data and our new eclipse timings in Fig.~\ref{fig:ocl}. The model is able to reproduce the data from the literature very well, however, it fails to fit our new data.

We check the dynamical stability of this model by
reproducing the same test performed by \citet{Beuermann2012}
using the \texttt{Mercury6}\footnote{We used the version available at
\url{https://github.com/4xxi/mercury}} \citep{Chambers1999} package.
We set the initial Keplerian parameters of the system with the binary as a single body of mass $M_\mathrm{bin} = M_1 + M_2$ at the centre of the system, as described in \citet{Beuermann2012},
and we use the same hybrid symplectic integrator. As a first test, we integrate for $10^4$ yr and, we find that the inner planet is ejected after $\sim 2500$~yr, in contrast with \citet{Beuermann2012}'s paper, who suggest that their proposed model is stable for $10^8$~yr.\par

We perform additional checks using the \texttt{radau} integrator within the \texttt{Mercury6} code, and also using the \texttt{python-C} package \texttt{rebound}\footnote{\url{https://rebound.readthedocs.io/en/latest/}} \citep{ReinLiu2012} with three of their different integrators, \texttt{ias15} \citep{ReinSpiegel2015}, \texttt{whfast} \citep{ReinTamayo2015} and \texttt{mercurius}.
All the simulations were run for $10^6$~yr, using a stepsize of 8.8 days (1/530 of $P_3$) with output every 308.9 days (1/15 of $P_3$).
Additionally, we test the stability with a new version of \texttt{Mercury6}, \texttt{Mercury6\_binary}\footnote{\url{https://github.com/rsmullen/mercury6_binary}}, a modified version of the original code by \cite{Smullen2016}, which allows to simulate both single and binary stars, treating the central star in the binary as a composite ``big body" instead of a single central object. Following the advice by the author, we use the \texttt{radau} integrator to perform the simulation, and we integrate for $10^6$~yr with the same step size described above. We consider a planet to escape or be ejected at a distance $> 150$~au.\par

\begin{figure}
	\centering
		\includegraphics[width=1.0\columnwidth]{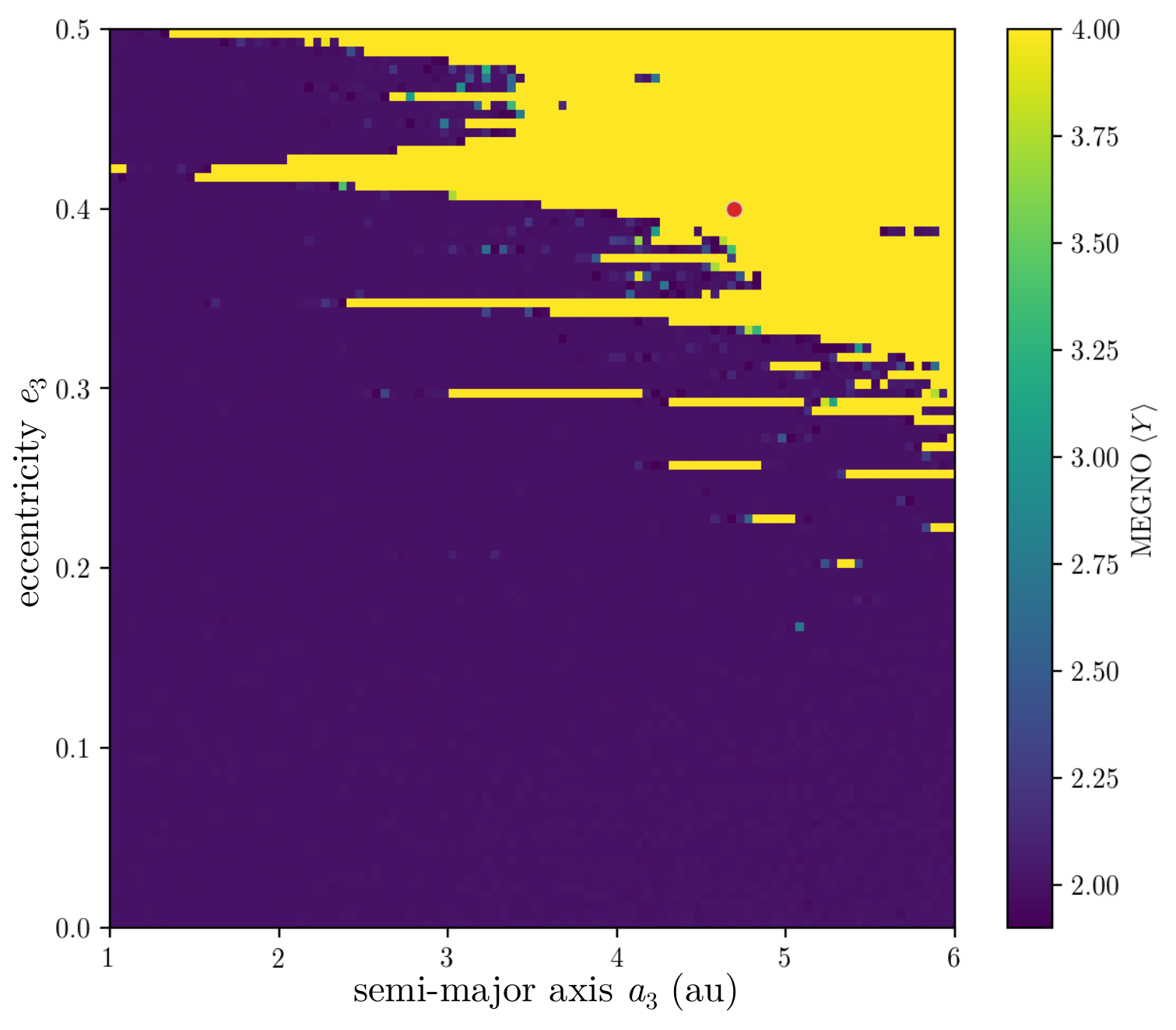}
	\caption[MEGNO simulations of previous model.]{
	MEGNO values, $\langle Y \rangle$, of each simulation based on \citet{Beuermann2012}'s solution with varying $a_3$ (1--6~au) and $e_3$ (0--0.5).
	To the simulations that did not complete the orbital integration or that returned $\langle Y \rangle > 4$, we assigned $\langle Y \rangle = 4$ (unstable).
	The configuration of \citet{Beuermann2012} is unstable and it is shown as the red dot (over-plotted on the yellow region).
	}
	\label{fig:megno}
\end{figure}

The initial orbital and physical parameters used for all the simulations performed are listed in Table~\ref{tab:sim}.
The results of all the simulations returned unstable systems, in different timescales and for different reasons, such as ejection of outer or inner planet, a close encounter between planets,
or the inner planet colliding with the binary.
As a final check, we use the Mean Exponential Growth factor of Nearby Orbits \citep[MEGNO,][]{CincottaSimo2000} indicator in \texttt{rebound}.
Briefly, the MEGNO indicator $\langle Y \rangle$, will reach the value of $\langle Y \rangle = 2$ for stable orbits, and it will be $\langle Y \rangle \gg 2$ for unstable configurations (in the case of $\langle Y \rangle > 4$ or a close encounter and an ejection,
we assign the maximum value $\langle Y \rangle = 4$).
We set the initial conditions as in Table~\ref{tab:sim}, but we let vary, for the inner companion (identified with the sub-index $3$),
the semi-major axis $a_3$ from 1 to 6~au and the eccentricity $e_3$ form 0 to $0.5$, both in 100 linear steps.
We compute the orbits of each configuration with the \texttt{whfast} integrator with a stepsize of 1 day for an integration time of $10^5$ years. The final grid has $100000$ simulations, each returning a MEGNO value.
As shown in Fig~\ref{fig:megno}, we find that the solution from \citet{Beuermann2012}, depicted by the red dot, lies on an unstable region, confirming our tests with different codes and integrators.
It is worth noting that all the simulations have the same reference frame as in \citet{Winn2010}, which is the plane $X$--$Y$ is the sky-plane and $\Omega_{3,4} = 180^\circ$, and we assume the orbits to be coplanar with the binary.

\begin{table}
	\centering
 	\caption{Orbital and physical parameters of HW Vir and the two companions proposed by \protect\cite{Beuermann2012} used for the dynamical stability tests, where the sub-indices bin, $3$ and $4$ represent the binary, the inner and outer companions, respectively.
 	 Values marked with * are assumed values.}
 	\label{tab:sim}
	\begin{tabular}{lc}
	\hline
    \hline
    	Parameter & Value \\
        \hline
        $M_\mathrm{bin}$ & 0.627 M\textsubscript{\astrosun} \\
      	$R_\mathrm{bin}$ & 0.860 R\textsubscript{\astrosun} \\
        $M_3$ &  $14.3\ M_\mathrm{J}$ \\
        $R_3$* & $1\ R_\mathrm{J}$ \\
        $a_3$ & $4.69$ au \\
        $e_3$ & $0.4$ \\
        $i_3$ & $80.9^{\circ}$ \\
        $\omega_3$ & $-18^{\circ}$ \\
        $\mathcal{M}_3$ & $33^{\circ}$ \\
        $\Omega_3$ & $180^{\circ}$ \\
        $M_4$ & $65\ M_\mathrm{J}$ \\
        $R_4$* & $2\ R_\mathrm{J}$ \\
        $a_4$ & $12.8$ au \\
        $e_4$ & $0.05$ \\
        $i_4$ & $80.9^{\circ}$ \\
        $\omega_4$ & $0^{\circ}$ \\
        $\mathcal{M}_4$ & $166.23^{\circ}$ \\
        $\Omega_4$ & $180^{\circ}$ \\
        \hline
	\end{tabular}
\end{table}

\section{A new model}
\label{sec:new}

Our aim at this stage is to find a new LTTE model that properly fits the data. We separately analyzed two data sets: one with all the available data (317 points), and one for which we discarded the first two observing seasons from the literature (35 photoelectric measurements between JD 2445730 and 2445745 from \citealt{Kilkenny1994}). From now on, we will refer to these data sets as the ``full'' and the ``reduced'' one, respectively.
The latter selection was done as a test
since the \citet{Kilkenny1994} data
were always suspiciously offset from any best-fit model and lack
the original time-series data, i.e., we are unable to perform any independent check on them.
We also rescale all the $T_0$ errors by adding in quadrature 1~s to \citet{Beuermann2012}'s and our values, and 5~s to the rest of the literature values.
We apply this rescaling to take into account systematic errors in the absolute calibration of the time stamps at this level (due for instance to clock drift, to the finite shutter travel time or to technical dead times while commanding the camera or saving the images). This assumption will be later empirically justified by the residual of our best-fitting models being very close to $\chi^2_r\simeq 1$.

After removing the outliers and rescaling the errors, we extend the code described in Section~\ref{sec:ltte} with the implementation of \texttt{PIKAIA} \citep{PIK_Charbonneau1995}, a genetic algorithm to solve multi-modal optimisation problems.
This algorithm is based on the theory of evolution by means of natural selection, that is, a new population is generated by choosing the fittest pairs from the original population, and this process continues until a certain fitness level is achieved or after a predefined number of generations.
We perform $100000$ simulations of 1000 generations each on a population of 200 individuals and we use the inverse of the reduced chi-square $1/\chi_\mathrm{r}^2$ as our fitness function.
Once the code computes the results for \texttt{PIKAIA} at the end of each simulation, it uses the Levenberg-Marquardt (LM) algorithm to refine the \texttt{PIKAIA} output and it calculates the final best-fitting solution.

We also run an independent analysis based on a modified version of \texttt{PIKAIA} in \texttt{Fortran 90}, wrapped in \texttt{python}, and coupled with the affine invariant ensemble sampler \citep{GoodmanWeare2010} algorithm implemented in the \texttt{emcee} package \citep{ForemanMackey2013}.
The \texttt{PIKAIA} part used 200 individuals (a set of parameters) for 2000 generations, while we run \texttt{emcee} with 100 walkers (or chains) for 10000 steps (we remove the initial 2000 steps as burn-in). We repeat this coupled analysis 1000 times.
\par

The same fitting parameters are used in both approaches, that is a linear ephemeris with reference time $T_\mathrm{ref}$ and period $P_\mathrm{bin}$, and the LTTE parameters for each $k$-th body,
i.e. $a_{k, \mathrm{bin}} \sin i$, period $P_k$, eccentricity $e_k$,
argument of pericentre $\omega_k$, and the time of the passage at pericentre $t_{\mathrm{peri},k}$.
We use the same boundaries of the fitting parameters for this code and the previous (see Table~\ref{tab:bounds}).
All the parameters have uniform-uninformative priors.\\

\begin{table}
	\centering
 	\caption{Boundaries of the parameters of linear ephemeris plus two LTTE model.}
 	\label{tab:bounds}
	\begin{tabular}{lcc}
	\hline
    \hline
    	Parameter & min & max\\
        \hline
        $T_\mathrm{ref}$ (BJD\textsubscript{TDB}) & 2445730.5 & 2445730.6 \\
      	$P_\mathrm{bin}$ (days) & 0.116719 & 0.116723 \\
        $a_{3,\mathrm{bin}}\sin i$ (au) & 0 & 1 \\
        $P_3$ (days) & 2000 & 10000 \\
        $e_3$ & 0 & 0.5 \\
        $\omega_3$~$(^\circ)$ & 0 & 360 \\
        $t_{\mathrm{peri},3}$ (days) & 2452000 & 2465000 \\
        $a_{4,\mathrm{bin}}\sin i$ (au) & 0.5 & 5 \\
        $P_4$ (days) & 10000 & 40000 \\
        $e_4$ & 0 & 0.7 \\
        $\omega_4$~$(^\circ)$ & 0 & 360 \\
        $t_{\mathrm{peri},4}$ (days) & 2452000 & 2491000 \\
        \hline
	\end{tabular}
\end{table}

We obtain a large set of solutions, but we select only the solutions that, first, are physically meaningful (i.e.~we discard negative eccentricity solutions, since LM is not bounded in the parameter intervals), and have a $\chi^2_\mathrm{r} < 2$. For each of these selected simulations, we run a stability\footnote{We compute the mass of the $k$-th companion combining the Third Kepler's law and $a_{k,\mathrm{bin}}=a_k M_k/(M_k + M_\mathrm{bin})$ and finding the real root of a polynomial of third order in $M_k$ of kind
$M_k^3 - x M_k^2 - 2 x M_\mathrm{bin}M_k - x M_\mathrm{bin}^2 = 0$ with $x = \frac{4 \pi^2}{G}  \frac{a_{k,\mathrm{bin}}^3}{P_k^2}$ and $k = 3\ \mathrm{and}\ 4$.} check with \texttt{rebound} and the MEGNO indicator.
We run simulations for $10^5$~yr with the \texttt{whfast} integrator and a small stepsize of 1 day.
We apply the full analysis (model fitting with two approaches and stability analysis) and find that all the solutions with $\chi^2_\mathrm{r} < 2$ are unstable for both data sets.
\par

\begin{figure*}
	\centering
		\includegraphics[width=1.0\textwidth]{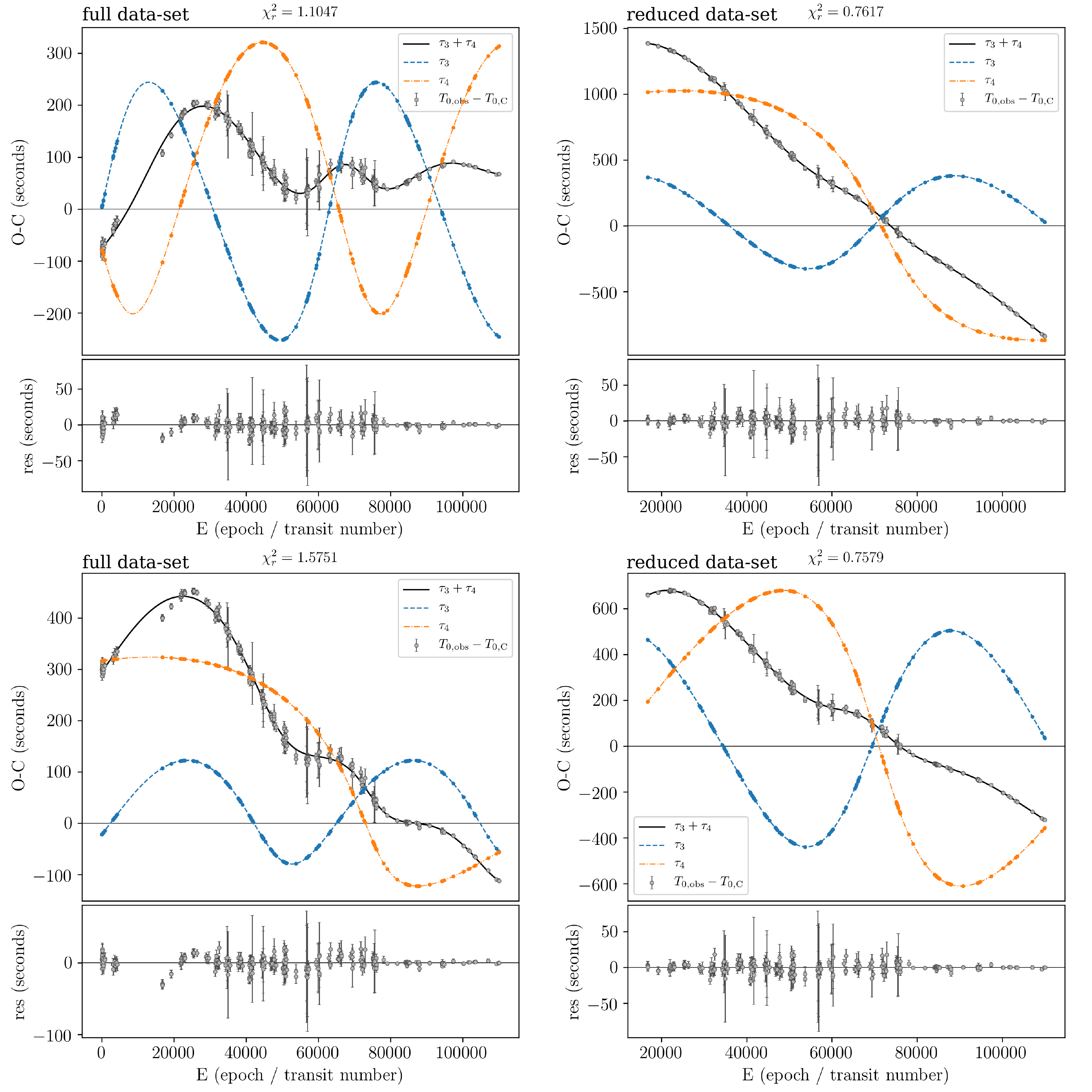}
	\caption[\textit{O-C} diagram for the two-companion model for HW Virginis]{The best two-companion models as the result of the fit to the full data set (left column) and the reduced data set (right column) from the best-fitting solution of the \texttt{PIKAIA+LM} (upper row) and of the \texttt{PIKAIA+emcee} (lower row) code.
	For each solution we show in the upper panel the $O-C$ (gray dots) as observed eclipse times ($T_{0,\mathrm{obs}}$) minus the linear ephemeris ($T_{0, \mathrm{c}}$), the combined LTTE of the two companions ($\tau_3 + \tau_4$ as black line), the single LTTE of the companions ($\tau_3$ and $\tau_4$ as blue dashed line and orange dash-dot line, respectively).
	The lower panel shows the residuals as $T_{0,\mathrm{obs}} - (T_{0, \mathrm{c}} + \tau_3 + \tau_4$).
	}
	\label{fig:ocmod}
\end{figure*}

We show in Fig.~\ref{fig:ocmod} the \textit{O-C} diagram for the two-companion model for the four best solutions (lowest $\chi_\mathrm{r}^2$) for both PIKAIA implementations and both data sets.
The four solutions show clearly different contributions from the inner (3) and outer (4) companions, with different periods, amplitudes, and patterns; yet, they fit the observed data points surprisingly well, especially on the ``reduced'' data set.
It is worth noting that both solutions on the full data set are not able to properly reproduce the general trend of the two observing seasons around epoch $20\,000$ (1989-1990), being forced to fit the earliest points by \citet{Kilkenny1994}.

\begin{table*}
	\centering
 	\caption{Orbital and physical parameters of our four best-fitting solutions for the ETVs of HW Vir with two-companion model.}
 	\label{tab:best}
	\begin{tabular}{lcccc}
	\hline
    \hline
        & \multicolumn{2}{c}{full data set} & \multicolumn{2}{c}{reduced data set}\\
    	Model and physical parameters & \texttt{PIKAIA+LM} & \texttt{PIKAIA+emcee} & \texttt{PIKAIA+LM} & \texttt{PIKAIA+emcee}\\
        \hline
        $T_\mathrm{ref}^{(a)}$ (\bjdtdb)      & $45\,730.557572$ & $45\,730.553198$ & $45\,730.538213$ & $45\,730.5492131$ \\
      	$P_\mathrm{bin}$ (days)               & $0.1167195$ & $0.1167196$ & $0.1167198$ & $0.1167196$ \\
        $a_{3,\mathrm{bin}}\sin i$ (au)       & $0.51$ & $0.20$ & $0.72$ & $0.96$ \\
        $P_3$ (days)                          & $7\,367$ & $7\,315$ & $8\,781$ & $8\,947$ \\
        $e_3$                                 & $0.235$ & $0.241$ & $0.159$ & $0.199$ \\
        $\omega_3$ (\degr)                    & $4$ & $242$ & $331$ & $340$ \\
        $t_{\mathrm{peri},3}^{(a)}$ (\bjdtdb) & $60\,499$ & $58\,757$ & $62\,135$ & $53\,506$ \\
        $a_3\sin i^{(b)}$ (au)                & $6.5$ & $6.4$ & $7.4$ & $7.6$ \\
        $M_3^{(b)}$ (M$_\mathrm{J}$)        & $56$ & $22$ & $70$ & $96$ \\
        $a_{4,\mathrm{bin}}\sin i$ (au)       & $0.53$ & $0.56$ & $2.58$ & $1.45$ \\
        $P_4$ (days)                          & $8\,012$ & $26\,155$ & $34\,258$ & $13\,649$ \\
        $e_4$                                 & $0.24$ & $0.7$ & $0.68$ & $0.445$ \\
        $\omega_4$ (\degr)                    & $251$ & $211$ & $185$ & $186$ \\
        $t_{\mathrm{peri},4}^{(a)}$ (\bjdtdb) & $70\,449$ & $54\,541$ & $54\,160$ & $54\,103$ \\
        $a_4\sin i^{(b)}$ (au)                & $6.9$ & $15$ & $18.6$ & $10.1$ \\
        $M_4^{(b)}$ (M$_\mathrm{J}$)        & $54$ & $26$ & $106$ & $110$ \\
        $\chi^2_\mathrm{r}$                   & $1.105$ & $1.575$ & $0.762$ & $0.758$ \\
        dof                                   & $258$ & $227$ & $258$ & $227$ \\
        \hline
	\end{tabular}
	\vspace{0mm}
    \begin{flushleft}
	$^{(a)}$: \bjdtdb{} $- 2\, 400\, 000$.\\
	$^{(b)}$: Physical parameter computed from the model parameters.\\
	\end{flushleft}
\end{table*}

In Table~\ref{tab:best} we present the orbital and physical parameters of these best-fitting solutions. Values for the masses of the companions are within the brown dwarf range.
We did not attempt to compute realistic errors (i. e., other than the nominal errors output from the LM fit) on the derived parameters due to the dynamical instability of all the solutions we found.

\begin{figure*}
	\centering
	\includegraphics[width=1.0\textwidth]{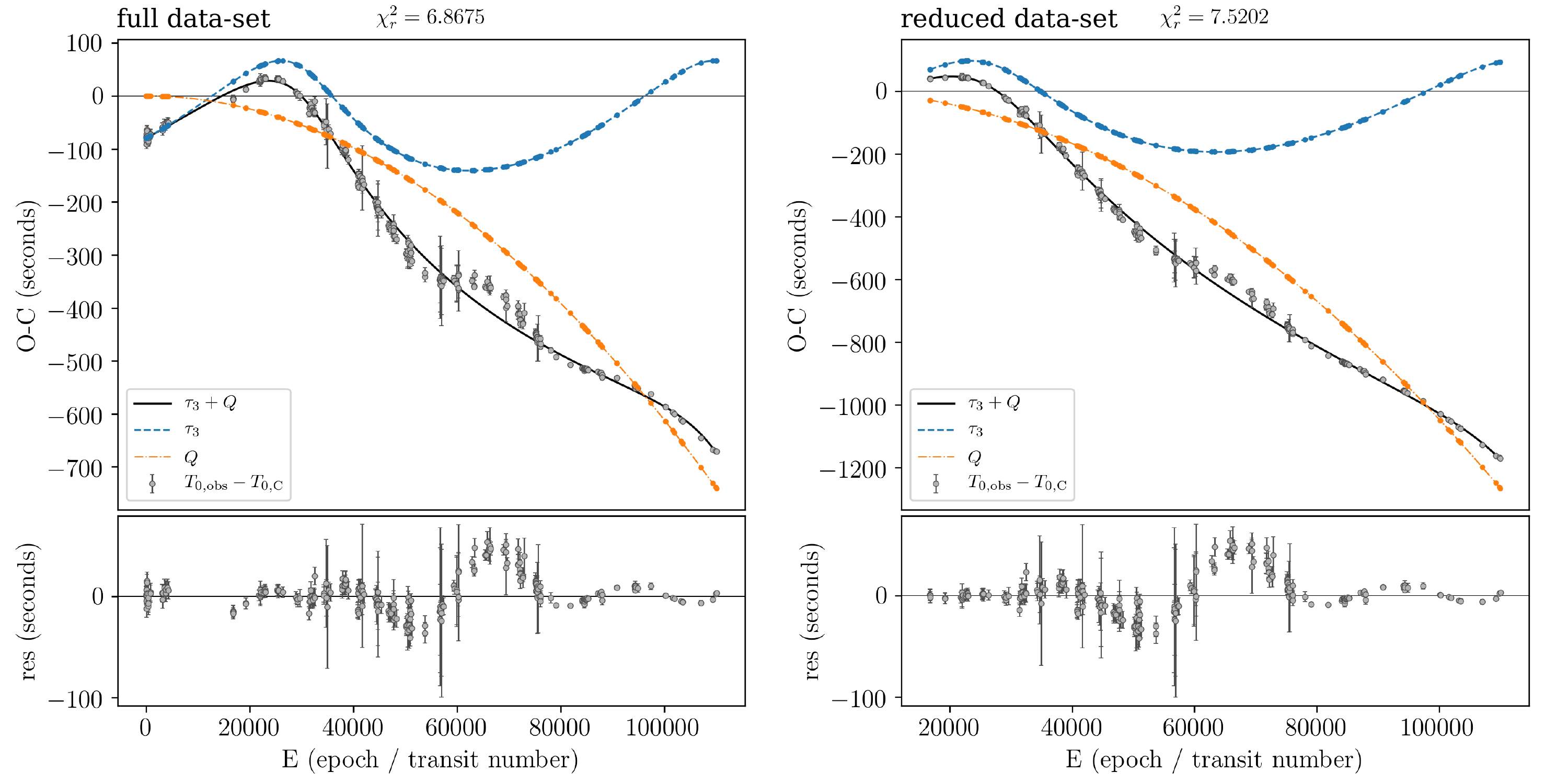}
	\caption[\textit{O-C} diagram for the one-companion model for HW Virginis]{The best one-companion models as the result of the fit to
	the full data set (left) and the reduced data set (right).
	Similar to Fig~\ref{fig:ocmod}, but now displaying the $Q$ term instead of $\tau_4$.
	The lower panel shows the residuals as $T_{0,\mathrm{obs}} - (T_{0, \mathrm{c}} + \tau_3 + Q$). Due to the high $\chi^2_\mathrm{r}$, these models are not suitable to explain the ETVs of HW Vir.
	}
	\label{fig:oc_1p}
\end{figure*}

Additionally, we test a different model with a linear ephemeris ($T_\mathrm{c}$), a one-companion LTTE ($\tau_3$), and a quadratic term ($Q$).
We apply this model to both data sets only with the \texttt{PIKAIA+emcee} approach.
We use uniform priors within the boundaries in Table~\ref{tab:1p+q}.
We find solutions with $\chi^2_\mathrm{r} > 6$  (see Table~\ref{tab:1p+q} and Fig~\ref{fig:oc_1p}) and BIC (Bayesian Information Criteria) that is higher than the two companion model,
for both the data sets. For this reason, we discard this model as a possible explanation for the ETVs.

\begin{table*}
	\centering
 	\caption{Boundaries and best-fitting parameters of the one companion model ($T_\mathrm{c} + \tau_3 + Q$).}
 	\label{tab:1p+q}
	\begin{tabular}{lcccc}
	\hline
    \hline
        & & & \multicolumn{2}{c}{best-fit} \\
    	Parameter & min & max & full data set & reduced data set\\
        \hline
        $T_\mathrm{ref}$ (\bjdtdb)      & $2\,445\,730.5$ & $2\,445\,730.6$ & $2\,445\,730.5575759$ & $2\,445\,730.5559335$\\
      	$P_\mathrm{bin}$ (days)         & $0.116719$ & $0.116723$ & $0.1167196$ & $0.1167197$ \\
        $a_{3,\mathrm{bin}}\sin i$ (au) & $0$ & $1$ & $0.213$ & $0.295$ \\
        $P_3$ (days)                    & $500$ & $50\,000$ & $9\,750$ & $10\,396$ \\
        $e_3$                           & $0$ & $0.5$ & $0.41$ & $0.37$ \\
        $\omega_3$~$(^\circ)$           & $0$ & $360$ & $123$ & $116$ \\
        $t_{\mathrm{peri},3}$ (days)    & $2\,452\,000$ & $2\,502\,000$ & $2\,459\,013$ & $2\,459\,294$ \\
        $Q$                             & $-10^{-8}$ & $10^{-8}$ & $-7.1\times 10^{-13}$ & $-1.2\times 10^{-12}$\\
        $\chi^2_\mathrm{r}$             & & & $6.868$ & $7.520$ \\
        dof                             & & & $262$ & $231$ \\
        \hline
	\end{tabular}
\end{table*}

\section{Discussion and conclusions}
\label{sec:con}

In this work we presented a study of the eclipsing binary system HW Vir by using hitherto unpublished photometric observations from four different facilities.
We converted all the light curve timings into a common reference frame, as it was crucial for the purposes of this work to have accurate and homogeneous time stamps in order to properly compare different data sets.
By combining our new timings with the ones available in the literature, we independently confirmed that the \citet{Beuermann2012} model reproduces the recent literature data until 2011, but it is unable to fit our new timings.
Additionally, we tested the dynamical stability of their proposed model and we found it to be unstable after only a few thousand years, opposite to their claim of $10^8$ yr of stability.\par

As a first effort to find a proper model for the LTTE in HW Vir, we used the \texttt{PIKAIA} code, which implements a genetic algorithm to explore the parameter space and estimate new parameters for the companions of the binary system.
We found a set of parameter vectors with a very good fit in a statistical sense, able to explain all the available data.
Notwithstanding, these sets of solutions led to very high values for
the masses of the companions of HW Vir ($\sim 50 M_{J}$, within the mass range of brown dwarfs) and dynamically unstable systems.

Regarding the recent work of \citet{Esmer2021}, we describe the most significant differences between their approach and ours in the following.
We performed a fully homogeneous analysis of all the new light curves presented, with the same tools and by fitting an accurate EB model (rather than measuring the $T_0$s with the \citealt{Kwee1956} method; \citealt{Li2018}). This, coupled with the use of larger telescopes, resulted in more accurate eclipse timings by a factor of five, on average. Also, we exploited a genetic algorithm to perform a comprehensive global search of the parameter space rather than a local one. For this reason, although our search for stable LTTE orbits has been unfruitful, the orbital parameters of our four new solutions fall well outside the region explored by \citet{Esmer2021}. The direct $O-C$ comparison of their $T_0$ with ours is also reassuring, as the average offsets of the residuals measured on a season-by-season basis demonstrates the sub-second accuracy in the absolute timestamp calibration of both data sets.

Although the best-fitting solutions we found were proven to be dynamically unstable, it is worth asking whether other stable orbital solutions with similar LTTE amplitudes exist, and how could we confirm or disprove them with one or more independent techniques.

The prospects for a follow-up with \emph{direct imaging} are not very promising in the short term.
The combination of angular separation (in our best solution, $0\overset{''}{.}11$ and $0\overset{''}{.}47$, respectively) and contrast ($\simeq 10^{-5}$ in the $K$ band if we assume the typical luminosity of a mature 50-$M_\mathrm{J}$ brown dwarf; \citealt{Phillips2020}) fall beyond or very close to the sensitivity limits of the existing ground based facilities such as SPHERE \citep{Beuzit2019} and GPI \citep{Ruffio2017}.
However, such systems may become very interesting targets for upcoming high-contrast imaging missions such as JWST and the Roman Space Telescope.

On the other hand, \emph{astrometry} as a follow-up approach could be much more feasible with the release in the near future of the individual astrometric measurements by GAIA \citep{Prusti2016}.
If we assume that the observed $O-C$ is entirely due to a combination of LTTE signals, its amplitude $A_{O-C}$ can be easily translated into the expected astrometric signal, $s$, as $s = A_{O-C} \times c / d$, where $d$ is the distance to HW Vir from Table~\ref{tab:hwv}.
We probe a range of $A_{O-C}$ from $100$ to $1500$ seconds, which is spanning the amplitude of the oscillating LTTE terms of the orbital solutions claimed in the recent literature and also compatible with those included in our two best-fitting models in Fig.~\ref{fig:ocmod}. We find that $s$ ranges from $1.10 \pm 0.12$~mas to $16.6 \pm 1.8$~mas for $A_{O-C} = 100$~s and $1500$~s, respectively.
That is in principle comfortably within the reach of GAIA sensitivity, since the expected astrometric precision of the individual positional measurements of HW Vir is $\sim 30$~$\mu$as \citep{Sahlmann2015}. In such scenario, the detection will be limited by the temporal baseline rather than the astrometric precision. Yet, if Gaia will survive up to its operational goal of ten years, at least the LTTE component with the shortest period can be robustly retrieved, while for the longest one a global analysis combining Gaia with the existing ETV data points will be needed.

A satisfying explanation for the ETVs of HW Vir is still eluding us, however, this only highlights the fact that there is still a lot to be learned about systems of this kind.
One of the challenges to accurately determine the underlying cause of the ETVs in this case, is that the observations show that the period of one of the components from the LTTE of HW Vir is longer than the total observational timespan available.
Therefore, increasing the observational baseline will certainly bring us closer to determine the cause behind the ETVs of HW Vir.\par

\section*{Acknowledgements}

GPi and LBo acknowledge the funding support from Italian Space Agency (ASI) regulated by ``Accordo ASI-INAF n. 2013-016-R.0 del 9 luglio 2013 e integrazione del 9 luglio 2015 CHEOPS Fasi A/B/C''.
L.T. acknowledges support from MIUR (PRIN 2017 grant 20179ZF5KS).
DNa acknowledges the support from the French Centre National d'Etudes Spatiales (CNES).


\section*{Data availability}

The data underlying this article will be uploaded on Vizier/CDS in a second stage, in the meantime, it will be shared on reasonable request to the corresponding author.



\bibliographystyle{mnras}
\bibliography{biblio}



\appendix

\section{Literature timings}

In this table we list the 240 timing measurements taken from the literature (from the compilation by \citealt{Kilkenny1994}, K94; \citealt{Lee2009}, L09; \citealt{Beuermann2012}, B12) and included in our fits together with our new data (Table~\ref{tab:t0}), after being converted by us in a uniform $\textrm{BJD}_\textrm{TDB}$ time standard \citep{Eastman2010}. The epoch is computed according to the ephemeris in Eq.~1 of \citep{Beuermann2012}.

\vspace{5mm}

\begin{supertabular}{llrl}
\hline\hline
$T_0$ ($\textrm{BJD}_\textrm{TDB}$) & $\sigma(T_0)$ & Epoch & Reference \\
\hline
  2445730.556669 & .000099 & 0 & \citetalias{Kilkenny1994}\\
  2445731.607139 & .000099 & 9 & \citetalias{Kilkenny1994}\\
  2445732.540889 & .000099 & 17 & \citetalias{Kilkenny1994}\\
  2445733.591389 & .000099 & 26 & \citetalias{Kilkenny1994}\\
  2445734.525149 & .000099 & 34 & \citetalias{Kilkenny1994}\\
  2445735.575549 & .000099 & 43 & \citetalias{Kilkenny1994}\\
  2445736.509219 & .000099 & 51 & \citetalias{Kilkenny1994}\\
  2445740.477899 & .000099 & 85 & \citetalias{Kilkenny1994}\\
  2445740.594559 & .000099 & 86 & \citetalias{Kilkenny1994}\\
  2445741.528339 & .000099 & 94 & \citetalias{Kilkenny1994}\\
  2445742.462240 & .000099 & 102 & \citetalias{Kilkenny1994}\\
  2445744.446450 & .000099 & 119 & \citetalias{Kilkenny1994}\\
  2445773.509431 & .000099 & 368 & \citetalias{Kilkenny1994}\\
  2445773.626191 & .000099 & 369 & \citetalias{Kilkenny1994}\\
  2445774.443131 & .000099 & 376 & \citetalias{Kilkenny1994}\\
  2445774.559881 & .000099 & 377 & \citetalias{Kilkenny1994}\\
  2445775.376921 & .000099 & 384 & \citetalias{Kilkenny1994}\\
  2445775.610421 & .000099 & 386 & \citetalias{Kilkenny1994}\\
  2445776.427511 & .000099 & 393 & \citetalias{Kilkenny1994}\\
  2445776.544181 & .000099 & 394 & \citetalias{Kilkenny1994}\\
  2445819.380354 & .000099 & 761 & \citetalias{Kilkenny1994}\\
  2445823.932404 & .000099 & 800 & \citetalias{Kilkenny1994}\\
  2446086.551616 & .000099 & 3050 & \citetalias{Kilkenny1994}\\
  2446098.573736 & .000099 & 3153 & \citetalias{Kilkenny1994}\\
  2446100.557976 & .000099 & 3170 & \citetalias{Kilkenny1994}\\
  2446101.608376 & .000099 & 3179 & \citetalias{Kilkenny1994}\\
  2446139.075518 & .000099 & 3500 & \citetalias{Kilkenny1994}\\
  2446164.403620 & .000099 & 3717 & \citetalias{Kilkenny1994}\\
  2446164.520380 & .000099 & 3718 & \citetalias{Kilkenny1994}\\
  2446203.271322 & .000099 & 4050 & \citetalias{Kilkenny1994}\\
  2446223.347073 & .000099 & 4222 & \citetalias{Kilkenny1994}\\
  2447684.326630 & .000065 & 16739 & \citetalias{Lee2009}\\
  2447687.244620 & .000065 & 16764 & \citetalias{Lee2009}\\
  2447688.295090 & .000076 & 16773 & \citetalias{Lee2009}\\
  2447689.228830 & .000065 & 16781 & \citetalias{Lee2009}\\
  2447968.539023 & .000061 & 19174 & \citetalias{Lee2009}\\
  2447972.507483 & .000061 & 19208 & \citetalias{Lee2009}\\
  2448267.574765 & .000061 & 21736 & \citetalias{Lee2009}\\
  2448294.887134 & .000099 & 21970 & \citetalias{Lee2009}\\
  2448295.003934 & .000099 & 21971 & \citetalias{Lee2009}\\
  2448295.937624 & .000099 & 21979 & \citetalias{Lee2009}\\
  2448307.609604 & .000061 & 22079 & \citetalias{Lee2009}\\
  2448311.578084 & .000061 & 22113 & \citetalias{Lee2009}\\
  2448313.562324 & .000061 & 22130 & \citetalias{Lee2009}\\
  2448365.385823 & .000059 & 22574 & \citetalias{Lee2009}\\
  2448371.455263 & .000059 & 22626 & \citetalias{Lee2009}\\
  2448404.370202 & .000059 & 22908 & \citetalias{Lee2009}\\
  2448406.354412 & .000065 & 22925 & \citetalias{Lee2009}\\
  2448410.322872 & .000061 & 22959 & \citetalias{Lee2009}\\
  2448682.512946 & .000061 & 25291 & \citetalias{Lee2009}\\
  2448684.497166 & .000059 & 25308 & \citetalias{Lee2009}\\
  2448703.522456 & .000076 & 25471 & \citetalias{Lee2009}\\
  2448704.456226 & .000059 & 25479 & \citetalias{Lee2009}\\
  2448705.506696 & .000059 & 25488 & \citetalias{Lee2009}\\
  2448803.317656 & .000059 & 26326 & \citetalias{Lee2009}\\
  2449104.453947 & .000065 & 28906 & \citetalias{Lee2009}\\
  2449122.312007 & .000061 & 29059 & \citetalias{Lee2009}\\
  2449137.368797 & .000061 & 29188 & \citetalias{Lee2009}\\
  2449139.353057 & .000059 & 29205 & \citetalias{Lee2009}\\
  2449190.242759 & .000099 & 29641 & \citetalias{Lee2009}\\
  2449393.567882 & .000076 & 31383 & \citetalias{Lee2009}\\
  2449400.571182 & .000076 & 31443 & \citetalias{Lee2009}\\
  2449418.546033 & .000099 & 31597 & \citetalias{Lee2009}\\
  2449427.533383 & .000076 & 31674 & \citetalias{Lee2009}\\
  2449437.571373 & .000099 & 31760 & \citetalias{Lee2009}\\
  2449450.643884 & .000099 & 31872 & \citetalias{Lee2009}\\
  2449476.322135 & .000099 & 32092 & \citetalias{Lee2009}\\
  2449480.407315 & .000099 & 32127 & \citetalias{Lee2009}\\
  2449485.309515 & .000099 & 32169 & \citetalias{Lee2009}\\
  2449511.337986 & .000099 & 32392 & \citetalias{Lee2009}\\
  2449518.341386 & .000099 & 32452 & \citetalias{Lee2009}\\
  2449519.274896 & .000099 & 32460 & \citetalias{Lee2009}\\
  2449728.552864 & .000099 & 34253 & \citetalias{Lee2009}\\
  2449733.571774 & .000099 & 34296 & \citetalias{Lee2009}\\
  2449778.625606 & .000503 & 34682 & \citetalias{Lee2009}\\
  2449785.628606 & .000208 & 34742 & \citetalias{Lee2009}\\
  2449808.505507 & .000702 & 34938 & \citetalias{Lee2009}\\
  2449833.483648 & .000099 & 35152 & \citetalias{Lee2009}\\
  2449880.288110 & .000099 & 35553 & \citetalias{Lee2009}\\
  2450142.556692 & .000099 & 37800 & \citetalias{Lee2009}\\
  2450144.540882 & .000099 & 37817 & \citetalias{Lee2009}\\
  2450147.575632 & .000099 & 37843 & \citetalias{Lee2009}\\
  2450155.512633 & .000091 & 37911 & \citetalias{Lee2009}\\
  2450185.392715 & .000099 & 38167 & \citetalias{Lee2009}\\
  2450186.443205 & .000099 & 38176 & \citetalias{Lee2009}\\
  2450201.383275 & .000099 & 38304 & \citetalias{Lee2009}\\
  2450202.433755 & .000099 & 38313 & \citetalias{Lee2009}\\
  2450216.673546 & .000099 & 38435 & \citetalias{Lee2009}\\
  2450218.424376 & .000099 & 38450 & \citetalias{Lee2009}\\
  2450222.509566 & .000099 & 38485 & \citetalias{Lee2009}\\
  2450280.285549 & .000099 & 38980 & \citetalias{Lee2009}\\
  2450491.430748 & .000083 & 40789 & \citetalias{Lee2009}\\
  2450491.547448 & .000076 & 40790 & \citetalias{Lee2009}\\
  2450506.487748 & .000099 & 40918 & \citetalias{Lee2009}\\
  2450509.522508 & .000099 & 40944 & \citetalias{Lee2009}\\
  2450510.572978 & .000099 & 40953 & \citetalias{Lee2009}\\
  2450511.506448 & .000070 & 40961 & \citetalias{Lee2009}\\
  2450511.506728 & .000099 & 40961 & \citetalias{Lee2009}\\
  2450543.721290 & .000099 & 41237 & \citetalias{Lee2009}\\
  2450547.456320 & .000099 & 41269 & \citetalias{Lee2009}\\
  2450547.689760 & .000099 & 41271 & \citetalias{Lee2009}\\
  2450552.475150 & .000208 & 41312 & \citetalias{Lee2009}\\
  2450575.468952 & .000099 & 41509 & \citetalias{Lee2009}\\
  2450594.377552 & .000702 & 41671 & \citetalias{Lee2009}\\
  2450595.427952 & .000099 & 41680 & \citetalias{Lee2009}\\
  2450596.361552 & .000099 & 41688 & \citetalias{Lee2009}\\
  2450597.295462 & .000099 & 41696 & \citetalias{Lee2009}\\
  2450599.279703 & .000099 & 41713 & \citetalias{Lee2009}\\
  2450600.330183 & .000099 & 41722 & \citetalias{Lee2009}\\
  2450631.260795 & .000099 & 41987 & \citetalias{Lee2009}\\
  2450883.491443 & .000099 & 44148 & \citetalias{Lee2009}\\
  2450885.475673 & .000099 & 44165 & \citetalias{Lee2009}\\
  2450910.453614 & .000099 & 44379 & \citetalias{Lee2009}\\
  2450912.321074 & .000208 & 44395 & \citetalias{Lee2009}\\
  2450912.554564 & .000099 & 44397 & \citetalias{Lee2009}\\
  2450927.494574 & .000208 & 44525 & \citetalias{Lee2009}\\
  2450931.346364 & .000099 & 44558 & \citetalias{Lee2009}\\
  2450943.368574 & .000116 & 44661 & \citetalias{Lee2009}\\
  2450943.485074 & .000208 & 44662 & \citetalias{Lee2009}\\
  2450946.403174 & .000503 & 44687 & \citetalias{Lee2009}\\
  2450948.387375 & .000603 & 44704 & \citetalias{Lee2009}\\
  2450955.390545 & .000099 & 44764 & \citetalias{Lee2009}\\
  2450959.242275 & .000099 & 44797 & \citetalias{Lee2009}\\
  2451021.220295 & .000099 & 45328 & \citetalias{Lee2009}\\
  2451183.576969 & .000099 & 46719 & \citetalias{Lee2009}\\
  2451190.580109 & .000099 & 46779 & \citetalias{Lee2009}\\
  2451216.491839 & .000099 & 47001 & \citetalias{Lee2009}\\
  2451236.567569 & .000099 & 47173 & \citetalias{Lee2009}\\
  2451300.413290 & .000208 & 47720 & \citetalias{Lee2009}\\
  2451301.346790 & .000116 & 47728 & \citetalias{Lee2009}\\
  2451301.463690 & .000116 & 47729 & \citetalias{Lee2009}\\
  2451302.397390 & .000208 & 47737 & \citetalias{Lee2009}\\
  2451326.324779 & .000099 & 47942 & \citetalias{Lee2009}\\
  2451368.227049 & .000099 & 48301 & \citetalias{Lee2009}\\
  2451578.555416 & .000099 & 50103 & \citetalias{Lee2009}\\
  2451582.523896 & .000099 & 50137 & \citetalias{Lee2009}\\
  2451608.552335 & .000099 & 50360 & \citetalias{Lee2009}\\
  2451616.489185 & .000208 & 50428 & \citetalias{Lee2009}\\
  2451627.460985 & .000345 & 50522 & \citetalias{Lee2009}\\
  2451630.145755 & .000070 & 50545 & \citetalias{Lee2009}\\
  2451630.262425 & .000070 & 50546 & \citetalias{Lee2009}\\
  2451654.423084 & .000208 & 50753 & \citetalias{Lee2009}\\
  2451655.356784 & .000208 & 50761 & \citetalias{Lee2009}\\
  2451668.429584 & .000099 & 50873 & \citetalias{Lee2009}\\
  2451671.463984 & .000099 & 50899 & \citetalias{Lee2009}\\
  2451674.382184 & .000208 & 50924 & \citetalias{Lee2009}\\
  2451688.038573 & .000076 & 51041 & \citetalias{Lee2009}\\
  2451689.088893 & .000124 & 51050 & \citetalias{Lee2009}\\
  2451691.423283 & .000116 & 51070 & \citetalias{Lee2009}\\
  2451692.356883 & .000116 & 51078 & \citetalias{Lee2009}\\
  2451712.315903 & .000099 & 51249 & \citetalias{Lee2009}\\
  2452001.429972 & .000116 & 53726 & \citetalias{Lee2009}\\
  2452001.546772 & .000116 & 53727 & \citetalias{Lee2009}\\
  2452342.251085 & .000059 & 56646 & \citetalias{Lee2009}\\
  2452348.437235 & .000712 & 56699 & \citetalias{Lee2009}\\
  2452348.553995 & .000902 & 56700 & \citetalias{Lee2009}\\
  2452349.487705 & .000099 & 56708 & \citetalias{Lee2009}\\
  2452353.456065 & .000404 & 56742 & \citetalias{Lee2009}\\
  2452356.490895 & .000099 & 56768 & \citetalias{Lee2009}\\
  2452373.298454 & .000722 & 56912 & \citetalias{Lee2009}\\
  2452373.415084 & .000872 & 56913 & \citetalias{Lee2009}\\
  2452402.361703 & .000099 & 57161 & \citetalias{Lee2009}\\
  2452410.298603 & .000099 & 57229 & \citetalias{Lee2009}\\
  2452431.308112 & .000099 & 57409 & \citetalias{Lee2009}\\
  2452650.390821 & .000061 & 59286 & \citetalias{Lee2009}\\
  2452675.368760 & .000065 & 59500 & \citetalias{Lee2009}\\
  2452724.390928 & .000394 & 59920 & \citetalias{Lee2009}\\
  2452724.507628 & .000394 & 59921 & \citetalias{Lee2009}\\
  2452756.371957 & .000070 & 60194 & \citetalias{Lee2009}\\
  2452759.406997 & .000523 & 60220 & \citetalias{Lee2009}\\
  2452764.425637 & .000503 & 60263 & \citetalias{Lee2009}\\
  2452764.542637 & .000208 & 60264 & \citetalias{Lee2009}\\
  2453061.360425 & .000065 & 62807 & \citetalias{Lee2009}\\
  2453112.716925 & .000059 & 63247 & \citetalias{Lee2009}\\
  2453112.833625 & .000061 & 63248 & \citetalias{Lee2009}\\
  2453124.972714 & .000107 & 63352 & \citetalias{Lee2009}\\
  2453360.746019 & .000059 & 65372 & \citetalias{Lee2009}\\
  2453384.323359 & .000059 & 65574 & \citetalias{Lee2009}\\
  2453410.702118 & .000116 & 65800 & \citetalias{Lee2009}\\
  2453444.083818 & .000116 & 66086 & \citetalias{Lee2009}\\
  2453444.200518 & .000116 & 66087 & \citetalias{Lee2009}\\
  2453465.443518 & .000208 & 66269 & \citetalias{Lee2009}\\
  2453466.377218 & .000116 & 66277 & \citetalias{Lee2009}\\
  2453491.355218 & .000076 & 66491 & \citetalias{Lee2009}\\
  2453773.933130 & .000059 & 68912 & \citetalias{Lee2009}\\
  2453825.289771 & .000061 & 69352 & \citetalias{Lee2009}\\
  2453829.024531 & .000306 & 69384 & \citetalias{Lee2009}\\
  2453829.141431 & .000208 & 69385 & \citetalias{Lee2009}\\
  2453861.589331 & .000059 & 69663 & \citetalias{Lee2009}\\
  2454105.182936 & .000116 & 71750 & \citetalias{Lee2009}\\
  2454108.217636 & .000116 & 71776 & \citetalias{Lee2009}\\
  2454108.334536 & .000116 & 71777 & \citetalias{Lee2009}\\
  2454143.233437 & .000116 & 72076 & \citetalias{Lee2009}\\
  2454143.350237 & .000116 & 72077 & \citetalias{Lee2009}\\
  2454155.255507 & .000059 & 72179 & \citetalias{Lee2009}\\
  2454155.372217 & .000059 & 72180 & \citetalias{Lee2009}\\
  2454158.290127 & .000091 & 72205 & \citetalias{Lee2009}\\
  2454214.082109 & .000065 & 72683 & \citetalias{Lee2009}\\
  2454216.416479 & .000116 & 72703 & \citetalias{Lee2009}\\
  2454239.410470 & .000208 & 72900 & \citetalias{Lee2009}\\
  2454498.877648 & .000116 & 75123 & \citetalias{Lee2009}\\
  2454498.877674 & .000060 & 75123 & \citetalias{Beuermann2012}\\
  2454509.148988 & .000065 & 75211 & \citetalias{Lee2009}\\
  2454509.265688 & .000059 & 75212 & \citetalias{Lee2009}\\
  2454512.300308 & .000059 & 75238 & \citetalias{Lee2009}\\
  2454513.350858 & .000083 & 75247 & \citetalias{Lee2009}\\
  2454514.167808 & .000059 & 75254 & \citetalias{Lee2009}\\
  2454514.284538 & .000059 & 75255 & \citetalias{Lee2009}\\
  2454515.335018 & .000065 & 75264 & \citetalias{Lee2009}\\
  2454517.319248 & .000059 & 75281 & \citetalias{Lee2009}\\
  2454533.193149 & .000116 & 75417 & \citetalias{Lee2009}\\
  2454533.309849 & .000116 & 75418 & \citetalias{Lee2009}\\
  2454535.177249 & .000404 & 75434 & \citetalias{Lee2009}\\
  2454554.902950 & .000503 & 75603 & \citetalias{Lee2009}\\
  2454588.401364 & .000070 & 75890 & \citetalias{Beuermann2012}\\
  2454601.707367 & .000060 & 76004 & \citetalias{Beuermann2012}\\
  2454607.076602 & .000065 & 76050 & \citetalias{Lee2009}\\
  2454608.593786 & .000061 & 76063 & \citetalias{Beuermann2012}\\
  2454611.628553 & .000059 & 76089 & \citetalias{Beuermann2012}\\
  2454841.916149 & .000059 & 78062 & \citetalias{Beuermann2012}\\
  2455543.984048 & .000014 & 84077 & \citetalias{Beuermann2012}\\
  2455549.003005 & .000014 & 84120 & \citetalias{Beuermann2012}\\
  2455556.006176 & .000015 & 84180 & \citetalias{Beuermann2012}\\
  2455582.968393 & .000015 & 84411 & \citetalias{Beuermann2012}\\
  2455584.952622 & .000015 & 84428 & \citetalias{Beuermann2012}\\
  2455591.955807 & .000015 & 84488 & \citetalias{Beuermann2012}\\
  2455593.006274 & .000014 & 84497 & \citetalias{Beuermann2012}\\
  2455605.028372 & .000014 & 84600 & \citetalias{Beuermann2012}\\
  2455605.962117 & .000019 & 84608 & \citetalias{Beuermann2012}\\
  2455615.883298 & .000014 & 84693 & \citetalias{Beuermann2012}\\
  2455635.725619 & .000013 & 84863 & \citetalias{Beuermann2012}\\
  2455647.864460 & .000014 & 84967 & \citetalias{Beuermann2012}\\
  2455648.914932 & .000014 & 84976 & \citetalias{Beuermann2012}\\
  2455654.750921 & .000013 & 85026 & \citetalias{Beuermann2012}\\
  2455680.779371 & .000014 & 85249 & \citetalias{Beuermann2012}\\
  2455682.763597 & .000019 & 85266 & \citetalias{Beuermann2012}\\
  2455896.010239 & .000014 & 87093 & \citetalias{Beuermann2012}\\
  2455953.903110 & .000021 & 87589 & \citetalias{Beuermann2012}\\
  2455957.988315 & .000014 & 87624 & \citetalias{Beuermann2012}\\
  2455977.013609 & .000014 & 87787 & \citetalias{Beuermann2012}\\
\hline
\end{supertabular}


\bsp	
\label{lastpage}
\end{document}